\documentclass{ws-procs9x6}

\def\barD{\overline D{}^0}

\def\barB{\overline B{}^0}
\def\Hbar{\overline{H}}

\def\beq{\begin{equation}}
\def\eeq{\end{equation}}
\def\bea{\begin{eqnarray}}
\def\eea{\end{eqnarray}}
\newcommand{\bra}[1]{\langle #1|}
\newcommand{\ket}[1]{|#1\rangle}

\begin{document}

\title{HEAVY MESON MOLECULES IN EFFECTIVE FIELD THEORY}

\author{MOHAMMAD T. ALFIKY, FABRIZIO GABBIANI, ALEXEY A. PETROV$^*$}

\address{Department of Physics and Astronomy, Wayne State University, \\
Detroit, Michigan, 48201, USA \\
$^*$Presenter. E-mail: apetrov@wayne.edu\\
www.physics.wayne.edu/\~{}apetrov/}

\begin{abstract}
We consider the implications from the possibility that the recently observed state
$X(3872)$ is a meson-antimeson molecule. We write an effective Lagrangian
consistent with the heavy-quark and chiral symmetries needed to describe $X(3872)$.
We explore the consequences of the assumption that $X(3872)$ is a
molecular bound state of $D^{*0}$ and $\barD$ mesons for the existence
of bound states in the $D^{0}\barD$ and $D^{*0}\overline D^{*0}$
\end{abstract}


\bodymatter

\section{Introduction}\label{intro}

The unusual properties of $X(3872)$ state, recently discovered in the 
decay $X(3872) \to J/\psi \pi^+\pi^-$, invited some speculations regarding 
its possible non-$c\bar c$ nature~\cite{Reviews,MolExp}. Since its mass lies
tantalizingly close to the $D^{*0}\barD$ threshold of 3871.3~MeV, it is
tempting to interpret $X(3872)$ as a $D^{*0}\barD$ molecule with $J^{PC}=1^{++}$ 
quantum numbers\cite{MoleculeX,Braaten}. 
Such molecular states can be studied using techniques of effective field 
theories (EFT). This study is possible due to the multitude of scales present in QCD.
The extreme smallness of the binding energy,
$E_b=(m^{\phantom{l}}_{D^0}+m^{\phantom{l}}_{D^{0*}})-M^{\phantom{l}}_X=
-0.6 \pm 1.1~\mbox{MeV}$,
suggests that this state can play the role of the ``deuteron''~\cite{MoleculeX}
in meson-antimeson interactions. This fact allows us to use methods similar to 
those developed for the description of the deuteron\cite{Weinberg,KSW}, with the added 
benefit of heavy-quark symmetry. A suitable effective Lagrangian describing such a 
system contains only heavy-meson degrees of freedom with interactions approximated by 
local four-boson terms constrained only by the symmetries of the theory. While the
predictive power of this approach is somewhat limited, several model-independent 
statements can be made. For instance, possible existence of a molecular state in 
$D^{*0}\barD$ channel does not imply a molecular state in $D^{*0}\overline {D^*}^0$ 
or $D^{0}\barD$ channels.

The tiny binding energy of this molecular state introduces an energy scale which is much 
smaller than the mass of the lightest particle, the pion, whose exchange can provide 
binding. This fact presents a problem with a straightforward EFT analysis of this 
bound state, which can be illustrated in the following example. Consider a low-energy 
s-wave scattering amplitude $A(k)$ of two heavy mesons with momentum $k$ 
(ignore spin for a moment),
\beq\label{ScatAmp}
A(k) = \frac{4\pi}{m_D} \frac{1}{k \cot\delta-ik}=
\frac{4\pi}{m_D} \frac{1}{\left(-1/a\right)+\left(r_0/2\right)k^2+...-ik},
\eeq
where $a$ is a scattering length, which is related to the binding energy $E_b$ of
a meson-antimeson bound state as $a \sim \left(E_b\right)^{-1/2}$. Naturally,
$a \sim 1/m$, where $m$ is the mass of exchanged particle that provides binding.
Scattering amplitude of Eq.~(\ref{ScatAmp}) can be obtained from an effective Lagrangian 
expressed in terms of only heavy-meson degrees of freedom as a power series in momentum $k$,
\beq\label{ExpScatAmp}
A(k) = \frac{4\pi}{m_D} \sum_n C_n k^n =
-\frac{4\pi a}{m_D} \left[
1 - iak + \frac{ar_0-2a^2}{2}k^2 + ...\right],
\eeq
where $C_n$ are the coefficients of that effective Lagrangian. A problem with a
simple application of EFT is apparent, as $a\to \infty$ for  $E_b \to 0$,
making the series convergence in Eq.~(\ref{ExpScatAmp}) problematic. Indeed,
$a \simeq 0.032$~MeV$^{-1}$ for X(3872), which is much larger than the inverse 
masses of possible exchange particles, $1/m_\pi \simeq 7.1 \times 10^{-3}$~MeV$^{-1}$,
$1/m_\rho \simeq 1.3 \times 10^{-3}$~MeV$^{-1}$, etc. This implies that all-order 
resummation of $(ak)^n$ series is required. In EFT language this would imply 
resummation of a class of ``bubble'' graphs, whose vertices are defined by $C_n$.

\section{The effective Lagrangian}\label{EL}

The general effective Lagrangian required for description of $D^{*0}\barD$ molecular state and
consistent with heavy-quark spin and chiral symmetries can be written as\cite{AlFiky:2005jd}
\beq\label{Lagr}
{\cal L}={\cal L}_1+{\cal L}_2,
\eeq
where the two-body piece ${\cal L}_2$ describes the interactions between heavy meson 
degrees of freedom. The one-body piece ${\cal L}_1$ describes strong interactions of
the heavy mesons $P$ and $P^*$ ($P=B,D$) containing one heavy quark $Q$ and
is well known\cite{Grinstein:1992qt}:
\bea\label{Lagr2}
{\cal L}_1 =-\mbox{Tr} \left[ \Hbar^{(Q)} \left(i v \cdot D 
+ \frac{D^2}{2 m^{\phantom{l}}_P} \right) H^{(Q)} \right] 
+ \frac{\lambda_2}{m^{\phantom{l}}_P}  \mbox{Tr}
\left[ \Hbar^{(Q)} \sigma^{\mu\nu } H^{(Q)} \sigma_{\mu\nu} \right] + ...
\eea
where the ellipsis denotes higher-order terms in chiral expansion, or describing 
pion-$H$ interactions and antimeson degrees of freedom $H_a^{(\overline{Q})}$ and 
$H_a^{(\overline{Q})\dagger}$.
A superfield describing the doublet of pseudoscalar heavy-meson 
fields $P_a = \left(P^0, P^+\right)$ and their vector counterparts with 
$v\cdot P^{*(Q)}_{a}=0$, is defined as 
$H_a^{(Q)}= (1/2)
(1+\not{v})\left[P^{*(Q)}_{a\mu}\gamma^\mu - P_a^{(Q)}\gamma_5\right]$. 
The second term in Eq.~(\ref{Lagr2}) accounts 
for the $P-P^*$ mass difference $\Delta\equiv m^{\phantom{l}}_{P^*}-m^{\phantom{l}}_P=
-2\lambda_2/m^{\phantom{l}}_P$. The two-body piece is\cite{AlFiky:2005jd}
\bea\label{Lagr4}
{\cal L}_2=&-&\frac{C_1}{4} \mbox{Tr} \left[ \Hbar^{(Q)} H^{(Q)} \gamma_\mu \right]
\mbox{Tr} \left[ H^{(\overline{Q})} \Hbar^{(\overline{Q})} \gamma^\mu \right]
\nonumber \\
&-&\frac{C_2}{4} \mbox{Tr} \left[ \Hbar^{(Q)}  H^{(Q)} \gamma_\mu \gamma_5 \right]
\mbox{Tr} \left[ H^{(\overline{Q})} \Hbar^{(\overline{Q})} \gamma^\mu \gamma_5 \right].
\eea
Heavy-quark spin symmetry implies that the same Lagrangian governs the four-boson
interactions of {\it all} $P_a^{(*)}=D^{(*)}$ states. Indeed, not all of these states 
are bound. Here we shall concentrate on $X(3872)$, which we assume to be a bound state of two 
{\it neutral} bosons, $P_a\equiv P^0\equiv D$\cite{MoleculeX}. Evaluating the traces 
yields for the $D\overline{D^*}$ sector
\bea\label{LocalLagr}
{\cal L}_{2,DD^*} = &-& C_1 D^{(c)\dagger} D^{(c)}
D^{*(\overline{c})\dagger}_\mu D^{* (\overline{c}) \mu}
- C_1 D^{*(c)\dagger}_\mu D^{*(c) \mu}
D^{(\overline{c})\dagger} D^{(\overline{c})} \nonumber \\
&+& C_2 D^{(c)\dagger} D^{*(c)}_\mu
D^{* (\overline{c})\dagger \mu} D^{(\overline{c})}
+ C_2 D^{* (c)\dagger}_\mu D^{(c)}
D^{(\overline{c})\dagger} D^{* (\overline{c}) \mu}
+\dots
\eea
As we show later, the resulting binding energy depends on 
a {\it linear combination} of $C_1$ and $C_2$. Similarly, one obtains the 
component Lagrangian governing the interactions of $D$ and $\overline D$,
\beq\label{LocalLagrPP}
{\cal L}_{2,DD} = C_1 D^{(c)\dagger} D^{(c)}
D^{(\overline{c})\dagger} D^{(\overline{c})}.
\eeq
Clearly, one cannot relate the existence of the bound state in the
$D\overline{D^*}$ and $D\overline{D}$ channels, as the properties of
the latter will depend only on $C_1$.

\section{Properties of bound states}

The lowest-energy bound state of $D$ and $\overline{D^*}$ is an eigenstate of 
charge conjugation,
\beq\label{Eigenstate}
\ket{X_{\pm}}=\frac{1}{\sqrt{2}}\left[
\ket{D^* \overline{D}} \pm \ket{D \overline{D}^*}
\right].
\eeq
To find the bound-state energy of $X(3872)$ with $J^{PC}=1^{++}$,
we shall look for a pole of the transition amplitude $T_{++}=\bra{X_+}T\ket{X_+}$.
Defining $DD^*$-$DD^*$ transition amplitudes,
\bea\label{Ts}
T_{11}&=&\langle D^* \overline{D}| T | D^* \overline{D} \rangle, \quad
T_{12}=\langle D^* \overline{D}| T | D \overline{D}^* \rangle,
\nonumber \\
T_{21}&=&\langle D \overline{D}^*| T | D^* \overline{D} \rangle, \quad
T_{22}=\langle D \overline{D}^*| T | D \overline{D}^* \rangle,
\eea
we also have to include a ``bubble'' resummation of loop contributions, as existence 
of a bound state is related to a breakdown of perturbative expansion\cite{Weinberg}. 
These amplitudes satisfy a system of Lippmann-Schwinger equations\cite{AlFiky:2005jd}.
In an algebraic matrix form,
\bea\label{LSEMatrix}
\left(
\begin{array}{c}
T_{11} \\
T_{12} \\
T_{21} \\
T_{22}
\end{array}
\right)
=
\left(
\begin{array}{c}
-C_1 \\
C_2 \\
C_2 \\
-C_1
\end{array}
\right)+
i\widetilde{A} \left(
\begin{array}{cccc}
-C_1 & C_2 & 0 & 0 \\
C_2 & -C_1 & 0 & 0 \\
0 & 0 & -C_1 & C_2 \\
0 & 0 & C_2 & -C_1
\end{array}
\right)
\left(
\begin{array}{c}
T_{11} \\
T_{12} \\
T_{21} \\
T_{22}
\end{array}
\right).
\eea
The solution of Eq.~(\ref{LSEMatrix}) produces the $T_{++}$ amplitude,
\beq\label{Solution}
T_{++}=\frac{1}{2}\left( T_{11}+T_{12}+T_{21}+T_{22} \right)=
\frac{\lambda}{1-i\lambda \widetilde{A}},
\eeq
where $\lambda=C_2-C_1$ and $\widetilde{A}$ is a (divergent) integral
\bea\label{Integral}
\widetilde{A}= \frac{i}{4} 2 \mu^{\phantom{l}}_{DD^*}
\int \frac{d^3q}{(2\pi)^3} \frac{1}{\vec{q}^{\;2}-2\mu^{\phantom{l}}_{DD^*}\left(E-\Delta\right)
-i\epsilon} 
\nonumber \\
= -\frac{1}{8 \pi} \mu^{\phantom{l}}_{DD^*}
|\vec{p}| \sqrt{1-\frac{2 \mu^{\phantom{l}}_{DD^*}\Delta}{\vec{p}^{\;2}}}.
\eea
Here $E=\vec{p}^2/2\mu^{\phantom{l}}_{DD^*}$, with $\mu^{\phantom{l}}_{DD^*}$ being the reduced 
mass of the $DD^*$ system. The divergence of the integral of Eq.~(\ref{Integral}) is removed 
by renormalization. We chose to define a renormalized $\lambda^{\phantom{l}}_R$ within the $MS$
subtraction scheme in dimensional regularization, which does not introduce any
new dimensionfull scales into the problem. In this scheme the integral $\widetilde{A}$ is
finite, which corresponds to an implicit subtraction of power divergences in
Eq.~(\ref{Integral}). This implies for the transition amplitude
\bea\label{FinAmp}
T_{++}=
\frac{\lambda^{\phantom{l}}_R}{1+(i/{8\pi})\lambda^{\phantom{l}}_R\, \mu^{\phantom{l}}_{DD^*}
|\vec{p}|
\sqrt{1-2 \mu^{\phantom{l}}_{DD^*}\Delta/{\vec{p}^{\;2}}}}.
\eea
The position of the pole of the molecular state on the energy scale
can be read off Eq.~(\ref{FinAmp}),
\beq\label{Pole}
E_{\rm Pole}=\frac{32 \pi^2}{\lambda_R^2 \mu_{DD^*}^3}-\Delta.
\eeq
Recalling the definition of binding energy $E_b$ and that 
$m^{\phantom{l}}_{D^*}$ = $m^{\phantom{l}}_{D}$ +  $\Delta$, we infer
\beq\label{Binding}
E_b=\frac{32 \pi^2}{\lambda_R^2 \mu_{DD^*}^3}.
\eeq
Assuming $E_b$ = 0.5 MeV, which is one sigma below the central value~\cite{MolExp}, 
and the experimental values for the masses~\cite{PDG}, we obtain 
$\lambda^{\phantom{l}}_R \simeq 8.4 \times 10^{-4} \ {\rm MeV}^{-2}$. 

Similar considerations apply to $D^0 \barD$ state, in which case the starting 
point is the Lagrangian term in Eq.~(\ref{LocalLagrPP}). Since it involves only a
single term, the calculations are actually easier and involve only
one Lippmann-Schwinger equation. The resulting binding energy is then\cite{AlFiky:2005jd}
\beq\label{BindingC}
E_b=\frac{256 \pi^2}{C_{1R}^2 m_D^3}.
\eeq
Examining Eq.~(\ref{BindingC}) we immediately notice that the existence of
a bound state in the $D^*\overline{D}$ channel does not dictate the properties
of a possible bound state in the $D^0 \barD$ or $B^0 \barB$ channels, since $C_1$ and
$C_2$ are generally not related to each other.
\begin{figure}[t]
\begin{center}
\psfig{file=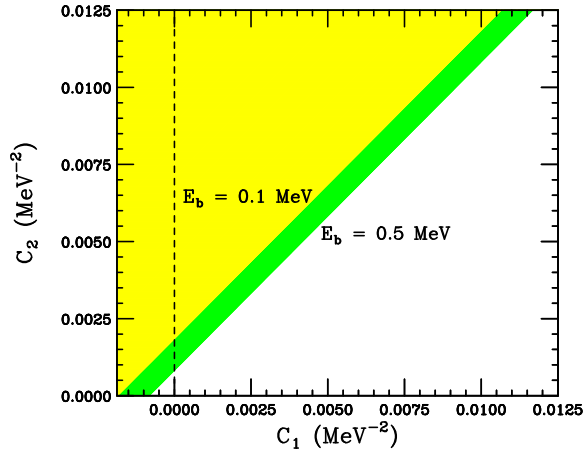,width=3.0in}
\end{center}
\caption{The coupling constant $C_2$ is plotted vs. $C_1$. The lightly shaded
area shows the region of parameter space allowed by postulating the
existence of a $J^{PC}$ = $1^{++}$ bound state with $E_b$ = 0.1 MeV,
while excluding the orthogonal bound state with C = --1. The darker
area becomes allowed in addition if we assume $E_b$ = 0.5 MeV.}
\label{fig2}
\end{figure}

If we assume that the orthogonal state with $J^{PC}=1^{+-}$ is not bound, which is
consistent with all the existing experimental observations, we can place some separate 
constraints on the renormalized values of $C_1$ and $C_2$. The amplitude orthogonal to $T_{++}$,
\beq\label{Ortho}
T_{--}=\bra{X_-}T\ket{X_-}=\frac{1}{2}\left( T_{11}-T_{12}+T_{21}-T_{22} \right)=
\frac{\lambda^{\prime}_R}{1-i\lambda^{\prime}_R \widetilde{A}_R},
\eeq
with $\lambda^{\prime}_R=-C_1-C_2$, does not have a pole that corresponds to
a bound state if $C_1+C_2 > 0$. The exclusion of the $C = -1$ state 
together with the assumption of the existence of the $C = +1$ state limits the ($C_1$,
$C_2$) parameter space as shown in Fig.~\ref{fig2}. 

\section{Conclusions}

We introduced an effective field theory approach in the analysis of the
likely molecular state $X(3872)$. We described its binding
interaction with contact terms in a heavy-quark symmetric
Lagrangian. The flexibility of this description allows us to ignore the
details of the interaction and to concentrate on its effects, namely a
shallow bound state and a large scattering length. We found that the 
existence of the bound state in the $D^*\overline{D}$ channel does not 
in general exclude a possibility of a bound state in the $D^0 \barD$ 
system, but does not require it. Future experimental studies of this state
are interesting\cite{ExpX} and should provide lots of new information about
properties of QCD bound states.

This work was supported in part by 
the U.S.\ Department of Energy under Contract DE-FG02-96ER41005. 
A.P. was also supported by the U.S.\ National Science Foundation
CAREER Award PHY--0547794.




\end{document}